\documentclass[fleqn,11pt]{article}
\usepackage{amsfonts,amssymb,epsfig,latexsym}

\newcommand{\Ga}{\alpha}
\newcommand{\Gb}{\beta}

\newcommand{\Gd}{\delta}

\newcommand{\GG}{\Gamma}

\newcommand{\Gl}{\lambda}

\newcommand{\Gs}{\sigma}

\newcommand{\Gth}{\theta}
\newcommand{\GTh}{\Theta}

\newcommand{\cA}{{\scriptscriptstyle\cal A}}
\newcommand{\cB}{{\scriptscriptstyle\cal B}}
\newcommand{\cC}{{\scriptscriptstyle\cal C}}
\newcommand{\cD}{{\scriptscriptstyle\cal D}}

\newcommand{\cM}{{\scriptscriptstyle\cal M}}
\newcommand{\cN}{{\scriptscriptstyle\cal N}}
\newcommand{\cK}{{\scriptscriptstyle\cal K}}
\newcommand{\cL}{{\scriptscriptstyle\cal L}}
\newcommand{\cP}{{\scriptscriptstyle\cal P}}
\newcommand{\cQ}{{\scriptscriptstyle\cal Q}}

\newcommand{\CD}{{\cal D}}

\newcommand{\CM}{{\cal M}}

\newcommand{\CL}{{\cal L}}

\newcommand{\CV}{{\cal V}}

\newcommand{\dA}{{\dot{A}}}
\newcommand{\dB}{{\dot{B}}}

\newcommand{\ft}[2]{{\textstyle {\frac{#1}{#2}} }}

\newcommand{\tr}{{\rm tr \,}}

\def\C{\mathbb{C}}

\def\R{\mathbb{R}}

\newcommand{\be}{\begin{equation}}
\newcommand{\ee}{\end{equation}}
\newcommand{\ben}{\begin{displaymath}}
\newcommand{\een}{\end{displaymath}}
\newcommand{\ba}{\begin{eqnarray}}
\newcommand{\ea}{\end{eqnarray}}

\newcommand{\non}{\nonumber\\}
\newcommand{\bea}{\begin{eqnarray}}
\newcommand{\eea}{\end{eqnarray}}
\newcommand{\bean}{\begin{eqnarray*}}
\newcommand{\eean}{\end{eqnarray*}}

\newcommand{\mathon}{\mathversion{bold}}
\newcommand{\mathoff}{\mathversion{normal}}

\def\moth{\mathsurround=0pt}
\newdimen\zo \zo=0pt

\def\tick{\leaders\hrule height 0.5ex depth 0pt \hskip 0.5pt}
\def\upboxfill{$\moth \setbox\zo\hbox{\tick}%
  \hskip 2pt\hbox to 0pt{$\tick$\hss}\hrulefill \hbox to 6pt{$\tick$\hss}$}
\def\underbox#1{\offinterlineskip{\mathord{\mathop{\vtop{\moth\ialign{##\crcr
      $\hfil\displaystyle{#1}\hfil$\crcr\noalign{}
      {\upboxfill}\crcr\noalign{}}}}\limits}}}
\def\dtick{\leaders\hrule height .34pt depth .5ex \hskip 0.5pt}
\def\downboxfill{$\moth \setbox\zo\hbox{\dtick}%
  \hskip 2pt\hbox to 0pt{$\dtick$\hss}\hrulefill \hbox to 6pt{$\dtick$\hss}$}

\def\undersym#1{\underbox{{}#1}}

\newcommand{\la}{\label}

\newcommand{\Ref}[1]{(\ref{#1})}

\newcommand{\maxl}{\omega}

\begin{document}
\thispagestyle{empty}

\begin{flushright}
AEI-2003-050 \\
ITP-UU-03/33 \\
SPIN-03/21
\end{flushright}

\bigskip

\begin{center}

\mathon
{\bf\Large Non-semisimple and Complex Gaugings}
\medskip

{\bf\Large of $N=16$ Supergravity}
\mathoff

\bigskip\bigskip\medskip

{\bf T. Fischbacher, H. Nicolai}
\vspace{.1cm}  

{\em Max-Planck-Institut f{\"u}r Gravitationsphysik,\\
  Albert-Einstein-Institut\\
  M\"uhlenberg 1, D-14476 Potsdam, Germany\\}
{\small tf@aei.mpg.de, 
nicolai@aei.mpg.de}

\vspace{.5cm}
{\bf H. Samtleben}
\vspace{.1cm}  

{\em Institute for Theoretical Physics} \,\&\, 
{\em Spinoza Institute},\\ 
{\em Utrecht University, Postbus 80.195, 3508 TD Utrecht, 
The Netherlands\\}
{\small h.samtleben@phys.uu.nl} 

\end{center}

\renewcommand{\thefootnote}{\arabic{footnote}}
\setcounter{footnote}{0}
\bigskip
\bigskip
\medskip

\begin{abstract}
Maximal and non-maximal supergravities in three dimensions allow 
for a large variety of semisimple (Chern-Simons) gauge groups. 
In this paper, we analyze non-semisimple and complex gauge groups 
that satisfy the pertinent consistency relations for a maximal 
($N=16$) gauged supergravity to exist. We give a general procedure 
how to generate non-semisimple gauge groups from known admissible 
semisimple gauge groups by a singular boost within $E_{8(8)}$. 
Examples include the theories 
with gauge group $SO(8)\times T_{28}$ that describe the reduction 
of IIA/IIB supergravity on the seven-sphere. In addition, we exhibit 
two `strange embeddings' of the complex gauge group $SO(8,\mathbb{C})$ 
into (real) $E_{8(8)}$ and prove that both can be consistently gauged. 
We discuss the structure of the associated scalar potentials as 
well as their relation to those of $D\geq 4$ gauged supergravities.
\end{abstract}

\renewcommand{\thefootnote}{\arabic{footnote}}
\setcounter{footnote}{0}

\newpage

\section{Introduction}
Locally supersymmetric theories in three space time dimensions have at
most $N\leq 16$ supersymmetries \cite{Juli83,MarSch83,dWToNi93}. For
$N>4$, the scalar sectors of these theories are governed by non-linear
$\Gs$-models over coset spaces $G/H$, where $H$ is always the maximal
compact subgroup of $G$ (the associated Lie algebras will be denoted
as $\mathfrak{g}\equiv {\rm Lie} \, G$ and $\mathfrak{h} \equiv {\rm
Lie}\, H$ throughout this paper). As shown only relatively recently,
these theories admit extensions where a subgroup $G_0\subset G$ is
promoted to a local symmetry \cite{NicSam00,NicSam01a,
NicSam01b,dWHeSa03}.  In contrast to higher dimensional gauged
supergravities, the vector fields in general appear via a Chern-Simons
(CS) rather than a Yang-Mills (YM) term. As it turns out, there is a
surprisingly rich structure and variety of possible gauge groups,
which have no analogs in higher dimensional ($D\geq 4$) gauged
supergravities. In particular, for the maximal $N=16$ theory, the
following semisimple subgroups of the global $E_{8(8)}$ symmetry have
been shown to be consistent gauge groups \cite{NicSam00,NicSam01a}
\ba\label{G0} 
G_0 &=& E_8 \;;\nonumber\\
G_0 &=& E_7 \times A_1 \;;\nonumber\\ 
G_0 &=& E_6 \times A_2 \;;\nonumber\\ 
G_0 &=& F_4 \times G_2 \;;\nonumber\\ 
G_0 &=& D_4 \times D_4 \;;
\ea 
and appear in all those real forms that `fit' into $E_{8(8)}$
(including, in particular, the groups $SO(p,q) \times SO(p,q)$ for
$p+q =8$). The situation is similar for lower~$N$ supergravities 
\cite{NicSam01b,dWHeSa03} where an equally rich variety
of gauge groups has been found to exist.

However, it is clear that the list \Ref{G0} cannot possibly
exhaust the groups for which a gauged maximal supergravity can be
constructed. First of all, it has been known for a long time that
in higher dimensions there exist gaugings with non-semisimple groups
\cite{Hull84a,Hull84b,Hull84c,ACFG00,ADFL02,Hull02,dWSaTr02},
implying that similar non-semisimple gaugings should also exist
in three dimensions. Secondly, the non-semisimple gaugings in 
three dimensions play a more prominent role than their higher 
dimensional cousins: as shown in~\cite{NicSam03a,dWHeSa03}, any 
three-dimensional YM gauged supergravity with gauge group $G_0$ 
is on-shell equivalent to a CS gauged supergravity with non-semisimple 
gauge group $G_0\ltimes T$ with a certain translation group $T$. 
This class in particular includes all theories obtained by 
reduction of higher dimensional maximal gauged supergravities
on a torus (i.e.\ a product of circles) or by Kaluza Klein 
compactification on some internal manifold (such as IIA/IIB
supergravity on the seven-sphere).

In this paper, we will identify these missing gauge groups, and in 
particular exhibit all those non-semisimple gaugings of the maximal 
$N=16$ theory which are equivalent on shell to the torus reductions of
the known gauged $N=8$ theories of \cite{dWiNic82,Hull84a,Hull84b,Hull84c} 
after performing the elimination of translational gauge fields 
described in \cite{NicSam03a}. However, due to the large number 
of possibilities we will not aim for an exhaustive classification 
of non-semisimple gaugings but rather study and explain in detail 
some representative examples of such gaugings. Our results
apply to lower $N<16$ gauged supergravities as well, furnishing
the $D=3$ supergravities associated with various supersymmetric
Kaluza Klein compactifications, and in particular the examples
listed in the last section of~\cite{NicSam03a}.

The second, and perhaps more surprising main result of the
present work is the admissibility of the {\em complex} gauge
group $SO(8,\C)$. The fact that the group $SO(8,\C)$ can be 
embedded into the {\em real} Lie group $E_{8(8)}$ (in two 
inequivalent ways) seems to have escaped mathematicians' notice 
so far. Because it does not require an imaginary unit, this 
embedding exhibits some rather strange properties, which we will 
highlight in section~4. Like the semisimple gauge groups \Ref{G0}, 
the $SO(8,\C)$ gauged supergravities cannot be derived from
higher dimensions by any known mechanism. Furthermore,
they feature a de Sitter stationary point at the origin breaking 
all supersymmetries, and with tachyonic instabilities. Similar complex
gaugings are expected to exist for lower $N<16$ supergravities
in three dimensions, in particular an $SO(6,\C)$ theory for $N=12$
and an $SO(5,\C)$ theory for $N=10$. We note that CS gauge theories 
with complex gauge groups are of considerable interest (\cite{Witt91}; 
see also \cite{Guko03} and references therein for some very recent
developments). The embedding of such theories into supergravity
with non-trivial matter couplings may well provide interesting
new perspectives.

We now list the main new gauge groups found in this paper for maximal 
$N=16$ supergravity, all of which are contained in $E_{8(8)}$
\ba\label{G1}
G_0 &=& SO(p,q) \ltimes T_{28} 
                \quad \mbox{ for $p+q=8$}\; ;  \nonumber\\
G_0 &=& CSO(p,q;r) \ltimes T_{p,q,r} 
        \quad \mbox{for  $p+q + r=8$ and  $r>0$} \; ; \nonumber\\
G_0 &=& SO(8,\C)
\;.
\ea
The first two of these are non-semisimple extensions of the 
groups $SO(p,q)$ and $CSO(p,q;r)$, respectively, and have not 
appeared in the supergravity literature before. Here, $T_{28}$ is 
an abelian group of 28 translations transforming in the adjoint 
of $SO(p,q)$. Similarly, $T_{p,q,r}$ is a group of translations, 
but of smaller dimension
\be
{\rm dim} \,  T_{p,q,r} = {\rm dim} \, CSO(p,q;r) = 28 - \ft12 r(r-1)
\;.
\ee
According to \cite{NicSam03a}, the elimination of the gauge fields 
associated with the translational subgroups produces a YM type 
gauged supergravity with YM gauge group $SO(p,q)$ or $CSO(p,q;r)$, 
respectively. The resulting YM gauged supergravities 
coincide with the ones that would be obtained by an $S^1$ reduction 
of the $SO(p,q)$ or $CSO(p,q;r)$ gauged $N=8$ supergravities in four 
dimensions \cite{Hull84a,Hull84b,Hull84c}. In addition to \Ref{G1}
we will exhibit some examples of purely nilpotent gaugings obtained 
by the boost method of section~2.

In \Ref{G1} the groups $SO(8)\ltimes T_{28}$ and $SO(8,\C)$ are singled 
out because they admit {\em two} inequivalent gaugings corresponding to 
the two embeddings
\be
\bf{248} = (\bf{28}, \bf{1}) \oplus  (\bf{1}, \bf{28}) \oplus
(\bf{8}_v, \bf{8}_v) \oplus  (\bf{8}_s, \bf{8}_s) \oplus (\bf{8}_c,
\bf{8}_c)
\;,
\la{decA}
\ee
(type IIA) and 
\be
\bf{248} = (\bf{28}, \bf{1}) \oplus  (\bf{1}, \bf{28}) \oplus
(\bf{8}_v, \bf{8}_v) \oplus  (\bf{8}_s, \bf{8}_c) \oplus (\bf{8}_c,
\bf{8}_s)
\;.
\la{decB}
\ee
(type IIB). The crucial feature
here is that only the compact real form $SO(8)$ admits 8-component 
{\em real} spinors and hence the phenomenon of triality. While the
compact $SO(8)\times SO(8)$ gaugings based on \Ref{decA} and \Ref{decB} 
are still equivalent, these two embeddings define different 
diagonal $SO(8)$ subgroups, leading to inequivalent embeddings 
and gaugings of $SO(8)\ltimes T_{28}$ and $SO(8,\C)$. 
Let us mention that these groups can be written uniformly as 
$SO(8,\C^\times)$, where $\C^\times$ is equal to the complex 
numbers~$\C$, the `split complex numbers'~$\C'$ or the 
`dual numbers'~$\C^0$
\cite{Rose97}
\ba
\C' &:=& \R \oplus \R e_1  \;, \nonumber\\
\C^0 &:=& \R \oplus \R e_0 \;,
\ea
with `imaginary units' obeying $e_1^2 = +1$ and $e_0^2 =0$, respectively,
because we have the following group isomorphisms \cite{Rose97}
\ba
SO(8,\C') &\cong& SO(8) \times SO(8) \;, \nonumber\\
SO(8,\C^0) &\cong& SO(8) \ltimes T_{28} \;.
\ea

By contrast, the other groups in \Ref{G1} involving $SO(p,q)$ or 
$CSO(p,q;r)$ (with $p\not=0,8$) admit {\em only one} embedding. 
This is obvious from the decomposition of the ${\bf 248}$ under 
the subgroup $SO(p,q)\times SO(p,q)\subset E_{8(8)}$
\be
\bf{248} = (\bf{28}, \bf{1}) \oplus  (\bf{1}, \bf{28}) \oplus
(\bf{8}_v, \bf{8}_v) \oplus  (\bf{8}_v, \bf{8}_v) \oplus (\bf{8}_v,
\bf{8}_v)
\;.
\ee
Alternatively, the existence of only one embedding follows from the fact 
that in the vacuum the symmetry is always broken to some compact subgroup 
of the gauge group involving factors of $SO(p)$ with $p < 8$, for which 
there is no triality.

\section{Generalities}

As shown in our previous work, the essential information about a 
maximally supersymmetric gauged $D=3$ supergravity is encoded in the 
so-called embedding tensor $\GTh$. This tensor characterizes the embedding 
of the gauge group into the global symmetry of the ungauged supergravity 
under consideration (see \cite{dWToNi93} for a classification of 
locally supersymmetric $\sigma$-models in three space time dimensions),
and allows us to immediately write down the Lagrangian and supersymmetry 
transformations from the formulas of \cite{NicSam01a} once $\GTh$ is 
explicitly given. For this reason and in order to keep this paper within 
reasonable proportions we will not present any explicit Lagrangians or 
supersymmetry transformations, but restrict attention to the embedding 
tensors and their properties. Readers are therefore advised to consult 
especially Refs. \cite{NicSam00,NicSam01a,NicSam03a} before delving 
into the details of the present article.

Here, we briefly summarize some basic properties of the embedding
tensor~$\Theta$ and the scalar field potential, and then explain 
the boost method, which allows us to derive many non-semisimple
gaugings from known (consistent) semisimple ones. This construction
works for any number $N$ of supersymmetries, and therefore the discussion
in that subsection will be kept general, whereas the rest of this paper 
is mostly devoted to the maximal $N=16$ theory.

\subsection{The embedding tensor}

Generally, gauging a subgroup $G_0\subset G$ corresponds to promoting 
the group $G_0$ to a local symmetry by making the minimal substitution
\be\label{gauging}
\partial_\mu \longrightarrow \partial_\mu + g \Theta_{\cM\cN} 
t^\cM B_\mu^{\;\cN} \;,
\ee
for any derivative acting on a field transforming under the global
symmetry $G$ (prior to gauging). Here, $g$ is the gauge coupling
constant, the fields $B_\mu^{\;\cM}$ are the
dual vector fields to the scalar fields, and before gauging transform
in the adjoint of the global symmetry group $G$. By $\{t^\cM\}$, we
denote a basis of $\mathfrak{g}\equiv{\rm Lie}\, G$ with 
\be
[ t^\cM , t^\cN ] = {f^{\cM\cN}}_\cP\, t^\cP  \;.
\la{fABC}
\ee 
The so-called embedding tensor \cite{NicSam00,NicSam01a}
\be\label{Th}
\Theta ~\equiv~ \Theta_{\cM\cN}\,t^\cM \otimes t^\cN \in
{\sf Sym}\,(\mathfrak{g}\otimes\mathfrak{g}) \;,
\ee
characterizes the embedding of the gauge group $G_0$ into the global 
symmetry group $G$, or more succinctly, the embedding of the associated 
Lie algebras $\mathfrak{g}_0\subset\mathfrak{g}$. A basis of the Lie algebra 
$\mathfrak{g}_0$ is then given by the generators $\GTh_{\cM\cN}t^\cN$. 
In particular, we have
\be
{\dim} \, \mathfrak{g}_0 = {\rm rank} \,\Theta \;.
\ee
Evidently, the components $\Theta_{\cM\cN}$ of the embedding tensor 
depend on the chosen basis and are thus defined only up to the 
adjoint action of $G$. They can thus assume various equivalent 
forms for a given gauge group $G_0$. 

To facilitate the task of writing out the components of a given 
embedding tensor, we will use the notation
\be
a\vee b := \ft12 (a \otimes b + b \otimes a) \;,
\ee
for the symmetric tensor product. We can also work with the dual 
embedding tensor $\GTh^{\cM\cN}\equiv \eta^{\cM\cK}\eta^{\cN\cL} 
\GTh_{\cK\cL}$, where indices are raised and lowered by means 
of the Cartan-Killing metric $\eta$ on the Lie algebra $\mathfrak{g}$, 
which we always assume to be non-degenerate (this requirement is 
evidently satisfied for the homogeneous target space manifolds 
appearing for $N>4$).

As shown in previous work~\cite{NicSam00,NicSam01a,dWHeSa03}, for a
consistent gauged supergravity to exist, the embedding tensor $\Theta$
must satisfy two conditions. First, the generators
$\GTh_{\cM\cN}t^\cN$ of the algebra $\mathfrak{g}_0$ must form a
closed algebra, under which $\GTh_{\cM\cN}$ is invariant, i.e.
\be\label{Theta3}
\Theta_{\cK\cP} \Theta_{\cL(\cM}\, {f^{\cK\cL}}_{\cN)} = 0 \;,
\ee
with the structure constants $f^{\cK\cL}{}_{\cN}$ from \Ref{fABC}.
Second, the embedding tensor needs to satisfy the projector condition
\be
\label{Theta1} {\mathbb{P}_{\cM\cN}}^{\cP\cQ} \,
\Theta_{\cP\cQ} = 0 \;,
\ee
where $\mathbb{P}$ projects onto the subrepresentation in
${\sf Sym}\,(\mathfrak{g}\otimes\mathfrak{g})$ which does {\em not} 
occur in the fermionic bilinears that can be built from the
gravitinos and the propagating fermions, see~\cite{dWHeSa03} for a
complete list (it is a non-trivial fact that
the $R$-symmetry representations arising from the fermionic bilinears
and compatible with local supersymmetry can always be assembled into
representations of the global symmetry group $G$). We call a subgroup 
of $G$ `admissible' if its embedding tensor obeys \Ref{Theta3} and 
\Ref{Theta1}, and hence gives rise to a consistent gauging.

Specializing to $N=16$ supergravity, the embedding tensor transforms
as an element of the tensor product
\be
\big({\bf 248} \otimes {\bf 248}\big)_{\rm sym} 
=  {\bf 1} \oplus{\bf 3875} \oplus {\bf 27000}
\;,
\ee
With the fermionic bilinears that can be built out of the gravitinos
and the matter fermions of $N=16$ supergravity equation \Ref{Theta1} becomes
\be\label{27000}
{\big(\mathbb{P}_{\bf 27000}\big)_{\cM\cN}}^{\cP\cQ} \,
\Theta_{\cP\cQ} = 0
\;.
\ee
Following \cite{MarSch83,KoNiSa00}, we split the generators of 
$\mathfrak{g}=\mathfrak{e}_{8(8)}$ into 120 compact ones 
$X^{IJ}= - X^{JI}$ with $SO(16)$ vector indices $I, J =1, \dots, 16$, 
and 128 noncompact ones $\{Y^A\}$ with $SO(16)$ spinor indices 
$A=1, \dots 128$. Then the condition \Ref{27000} implies that only 
particular $SO(16)$ representations can appear in $\GTh$: we have
\be
\Theta = \GTh_{IJ|KL}\, X^{IJ} \vee X^{KL} + 
         2 \GTh_{IJ|A} \,X^{IJ} \vee Y^A + \GTh_{A|B}\, Y^A \vee
Y^B
\;,
\ee 
with \cite{NicSam01a,FiNiSa02}
\ba\la{Theta}
\GTh_{IJ|KL} &=& -2 \Gth \Gd^{IJ}_{KL} +
  2 \Gd\undersym{_{I[K} \, \Xi_{L]J} \, } + \Xi_{IJKL}  \;, \non
\GTh_{IJ|A}&=& - \ft17 \GG^{[I}_{A\dA} \Xi^{J]\dA}   \;, \non
\GTh_{A|B} &=& \Gth \Gd_{AB} + \ft1{96} \Xi_{IJKL} \GG^{IJKL}_{AB}
\;,
\ea
and the $SO(16)$ $\GG$ matrices $\GG^I_{A\dA}$, where the indices
$\dA=1, \dots, 128$ label the conjugate spinor representation. 
The tensors $\Xi_{IJ}$, $\Xi_{IJKL}$ and $\Xi^{I\dA}$ transform as
the $\bf{135}$, $\bf{1820}$ and $\bf\overline{1920}$ representations 
of $SO(16)$, respectively; hence $\Xi_{II} = 0=\GG^I_{A\dA}
\Xi^{I\dA}$, and $\Xi_{IJKL}$ is completely antisymmetric in its 
four indices. Unlike for the semisimple gauge groups \Ref{G0} the singlet 
contribution in \Ref{Theta} is absent for non-semisimple and complex 
gauge groups, and we will thus set $\Gth = 0$ in the remainder.

For semisimple gaugings, the Lie algebra $\mathfrak{g}_0$
decomposes as a direct sum
\be\label{sumg}
\mathfrak{g}_0 = \bigoplus_i \mathfrak{g}_{0i}
\;,
\ee
of simple Lie algebras $\mathfrak{g}_{0i}$. The embedding tensor 
can be written as a sum of projection operators 
\be\label{projector}
\eta^{\cM\cP}\Theta_{\cP\cN} = \sum_i \varepsilon_i 
\, {(\Pi_i)^\cM}_\cN \;,
\ee
where $\Pi_i$ projects onto the $i$-th simple factor $\mathfrak{g}_{0i}$, 
and the constants $\varepsilon_i$ characterize the relative strengths
of the gauge couplings. There is only one overall gauge coupling 
constant $g$ for the maximal theory ($N=16$), but there may be several 
independent coupling constants for lower $N$. For the semisimple examples
with maximal supersymmetry known up to now \cite{NicSam00,NicSam01a}, 
the sum~\Ref{projector} contains at most two terms. Moreover, 
for all these gaugings we have
\be
\Xi^{I\dA}= 0 \qquad \mbox{(for semisimple $\mathfrak{g}_0$)}
\;.
\ee
As shown in~\cite{NicSam01a,FiNiSa02}, this implies that all these 
theories possess maximally supersymmetric (AdS or Minkowski) ground
states.

For non-semisimple gaugings, \Ref{sumg} is replaced by
\be\label{sumg1}
\mathfrak{g}_0 =  \bigoplus_i \mathfrak{g}_{0i} \oplus \mathfrak{t}
\;,
\ee
where $\mathfrak{t}$ represents the solvable part of the gauge group. 
As we will see below, for the non-semisimple gauge groups which
appear in our analysis, the latter subalgebra decomposes into
\be
\mathfrak{t} = \mathfrak{t}_0 \oplus \mathfrak{t}'
\;,
\ee
where $\mathfrak{t}_0$ transforms in the adjoint of the semisimple
part of the gauge group and pairs up with the semisimple subalgebra
in the embedding tensor, which has non-vanishing components 
only in $\mathfrak{g}_{0i}\vee \mathfrak{t}_0$ and in
$\mathfrak{t}'\vee\mathfrak{t}'$. We will also encounter examples 
of purely nilpotent gaugings, where the semi-simple part is absent. 

For the non-semisimple gaugings, in general all components in \Ref{Theta} 
are non-vanishing, in particular
\be
\Xi^{I\dA}\neq 0 \qquad \mbox{(for non-semisimple $\mathfrak{g}_0$)}
\;.
\ee
It is evident that \Ref{projector} cannot be valid for 
non-semisimple gauge groups because the Cartan-Killing metric
degenerates on the nilpotent part of the associated Lie
algebra. Furthermore, the complex gauge group $SO(8,\C)$ whose
admissibility we shall demonstrate here, also fails to satisfy \Ref{sumg}
when written in the real basis of $E_{8(8)}$; in fact,
\be
\Gth = \Xi_{IJ}=\Xi_{IJKL} = 0 \;, \qquad 
\mbox{(for $\mathfrak{g}_0 = \mathfrak{so}(8,\C)$)} \;,
\ee
so that $\Xi^{I\dA}$ represents its {\em only} nonvanishing component 
in \Ref{Theta}. The associated ground state is of de Sitter type and 
supersymmetry is completely broken, see section~4.

We finally note that the consistency conditions \Ref{Theta3},
\Ref{Theta1} remain covariant under the {\em complexified} global
symmetry group $E_8(\C)$. Indeed, non-semisimple gaugings in four
dimensions were originally found in~\cite{Hull84a,Hull84b,Hull84c}
by analytic continuation of $SO(8)$ in the complexified global 
symmetry group $E_7(\C)$. In three dimensions, a similar construction 
should exist relating the different non-compact real forms of the 
gauge groups \Ref{G0}, and explaining why the ratios of coupling 
constants between the two factor groups are the same independently of 
the chosen real form. Likewise, the gauge groups $SO(8)\times SO(8)$ 
and $SO(8,\C)$ are presumably related by analytic continuation 
in $E_8(\C)$. We will, however, present a more systematic and 
more direct construction based on an analysis of the consistency 
conditions~\Ref{Theta3}, \Ref{Theta1}.

\subsection{Some properties of the scalar potential}

The embedding tensor $\GTh_{\cM\cN}$ discussed in the previous section
completely specifies the gauged supergravity, i.e. its Lagrangian and
supersymmetry transformation rules. For the reader's convenience, and
because we will refer to them later, we here briefly recall some
pertinent formulas for the $N=16$ theory from \cite{NicSam00,NicSam01a}. 
Both the fermionic mass tensors and the scalar potential may be
expressed in terms of the so-called $T$-tensor
\be\label{Ttensor}
T_{\cA\cB} = \CV^\cM_{\;\; \cA} \CV^\cN_{\;\;\cB} \Theta_{\cM\cN}
\;,
\ee
where $\CV^\cM_{\;\; \cA}\in E_{8(8)}$ is a group valued matrix (the
248-bein) that combines the scalar fields of the theory. The fermions
in the theory are the 16 gravitini $\psi_\mu^I$ and the 128 spin-$1/2$
matter fermions $\chi^\dA$. They arise in the Lagrangian in bilinear
combinations contracted with the scalar (Yukawa) tensors
\ba
A_{1}^{IJ} &\equiv& 
\ft87\,\Gth\,\Gd_{IJ}
+\ft1{7}\,T_{{IK},{JK}}
\;,
\non[1ex]
A_{2}^{I\dA}&\equiv&
-\ft17\,\GG^J_{A\dA}\,T_{{IJ},{A}}
\;,
\non[1ex]
A_{3}^{\dA\dB}&\equiv&
2\Gth\,\Gd_{\dA\dB}
+\ft1{48}\,\GG^{IJKL}_{\dA\dB}\,
T_{{IJ},{KL}}
\;.
\la{AinT}
\ea
The scalar potential $W$ of maximal $N=16$ supergravity has a rather 
simple form in terms of the $T$-tensor~\Ref{Ttensor}, but becomes 
an extremely complicated function when expressed directly in terms 
of the 128 physical scalar fields \cite{Fisc02,Fisc03}. It reads
\ba\la{potential}
W &\equiv&
-\ft18 \,g^2\,\Big(
A_{1}^{IJ}A_{1}^{IJ}-\ft12\,A_{2}^{I\dA}A_{2}^{I\dA} \Big) \;.
\ea
In \cite{NicSam01a} it is shown that the extrema of the potential 
must obey the (necessary and sufficient) condition
\ba
3\,A_1^{IM}A_2^{M\dA} - A_3^{\dA\dB}A_2^{I\dB} 
&\stackrel{!}{=}& 0
\;,
\la{stationary}
\ea
This condition is met in particular if $A_2^{I\dA} =0$, as is the
case for all semisimple gaugings in \Ref{G0} at the trivial stationary 
points $\CV = {\bf 1}$. Moreover, the vanishing of $A_2^{I\dA}$ implies
maximal supersymmetry of these groundstates, which are therefore stable
by the general analysis of \cite{GiHuWa83}. As we will see, the complex
gauge group $SO(8,\C)$ realizes another possibility to satisfy
\Ref{stationary}: there we have $A_{1}^{IJ}=0$ and $A_3^{\dA\dB}=0$ 
for $\CV={\bf 1}$.

\subsection{The boost method}

To find gaugings with non-semisimple groups in three dimensions, 
one can either directly search for solutions of the above two 
conditions~\Ref{Theta3}, \Ref{Theta1}, or try to generate new
solutions to these equations from
known semisimple ones. A convenient alternative method realizing this
possibility is the `boost method', which we will now explain.

The method can be applied to any admissible 
semisimple gauge group $G_0\subset G$. Having chosen a suitable $G_0$,
one selects a non-compact (`boost') generator $N\in\mathfrak{g}$, such 
that $N\notin \mathfrak{g}_0$. This boost generator will 
preserve a (still semisimple) subgroup $\tilde G_0 \subset G_0$, i.e.
$[N,\tilde\mathfrak{g}_0]=0$ where $\tilde\mathfrak{g}_0$ is the
associated stable subalgebra of $\mathfrak{g}_0$. 
The deformation can be understood
systematically by decomposing (`grading') the full Lie algebra 
$\mathfrak{g}$ into eigenspaces of $N$ under its adjoint action. 
That is,
\be
\mathfrak{g} = \bigoplus_{j=-\ell}^{\ell} \mathfrak{g}^{(j)}
\;,
\label{grading}
\ee
where $ [N, t] = j t$ for $t\in\mathfrak{g}^{(j)}$ and 
$\ell$ is the maximum eigenvalue under the adjoint action of $N$.
Obviously $\tilde\mathfrak{g}_0 =\mathfrak{g}^{(0)} \cap
\mathfrak{g}_0$.

Since the embedding tensor $\Theta$ has two indices in the adjoint 
representation of $\mathfrak{g}$, we have a similar decomposition
for it, viz.
\be\label{gradeTheta}
\Theta = \sum_{j=-2\ell}^{2\ell} \Theta^{(j)}
\;.
\ee
Under the action of the boost $\exp(\lambda N)$ the embedding tensor 
scales as
\be
\exp(\lambda N): \Theta 
\longrightarrow \sum_{j=-2\ell}^{2\ell} e^{j\lambda}\Theta^{(j)}
\;.
\label{boostTH}
\ee
The graded pieces $\Theta^{(j)}$ themselves need not transform
irreducibly under $\tilde\mathfrak{g}_0$. We now exploit two basic 
properties of $\Theta$ and the consistency conditions 
\Ref{Theta3},\Ref{Theta1}, namely 
\begin{itemize}
\item the covariance of $\Theta$ w.r.t.\ to the global symmetry
group $G$, 
implying that a `rotated' embedding tensor still satisfies the 
conditions \Ref{Theta3} and \Ref{Theta1}, and
\item the fact that these conditions remain valid under
rescaling of $\GTh$. 
\end{itemize}
We thus consider the boosted embedding tensor~\Ref{boostTH}
and simultaneously replace $g \rightarrow g e^{-\maxl\lambda}$ for the
gauge coupling constant multiplying $\GTh$ in~\Ref{gauging},
 where $\maxl$ is the maximum degree appearing in the
decomposition \Ref{gradeTheta}
of $\Theta$ (and might be different from $2\ell$). While the
resulting
$\Theta$ is still equivalent to the original one for any finite $\Gl$, 
this needs no longer be true for the limit
$\lambda\rightarrow\infty$.
By continuity, the new embedding tensor 
\be\label{Thetaboost1}
\overline\Theta := \lim_{\lambda\rightarrow\infty} 
\left( e^{-\maxl\Gl}
        \exp(\lambda N) (\Theta)\right) \;,
\ee
still satisfies the projector condition \Ref{Theta1} and the quadratic 
condition \Ref{Theta3}. Given a particular 
grading as in \Ref{gradeTheta}, it is now easy to see that only 
the highest components survive in this limit, viz.
\be\label{Thetaboost2}
\overline\Theta^{(j)} := \left\{ \begin{array}{ll}
          \Theta^{(j)} & \mbox{for $j = \maxl$} \\
               0  & \mbox{otherwise}
                      \end{array}
                      \right.
                      \;.
\ee
We emphasize once more that the graded piece $\overline\Theta{}^{(j)}$ 
may have more than one irreducible component. Furthermore, it is
easy to see that the limiting gauge group will be solvable unless
the graded components $\overline\Theta{}^{(j)}$ contain a piece
intersecting the compact subalgebra.

While the structure constants $f$ of 
the global symmetry group are not affected by the boost because they 
are invariant w.r.t.\ to the global group $G$, the boosted structure
constants of the gauge group will no longer be equivalent to the
original ones.\footnote{The limit \Ref{Thetaboost1} therefore
realizes the well known Wigner-In\"on\"u contraction.} If the 
embedding tensor allows more than one 
free gauge coupling constant, we have the freedom to also scale 
the independent gauge couplings independently in such a way
that different limits $\lambda\rightarrow\infty$ give rise to
inequivalent new solutions of \Ref{Theta3}, \Ref{Theta1}.

For each ``seed'' gauge group $G_0$ the non-semisimple gauge groups 
that can be generated by this method can be systematically 
searched for by ($i$) identifying an appropriate boost generator $N$, 
and ($ii$) decomposing the embedding tensor $\Theta$ into graded pieces 
according to \Ref{gradeTheta}. The problem can therefore be reduced to 
the classification of all possible graded decompositions of the Lie algebra 
$\mathfrak{g}$, and to analyzing how $\Theta$ decomposes under them. 
The first problem, in turn, can be reformulated in terms of graded 
decompositions of the associated root systems.

An equally important consequence of the above derivation is that the
projector condition \Ref{Theta1} is in fact satisfied grade by grade
in the decomposition~\Ref{gradeTheta}. Given any embedding tensor
satisfying \Ref{Theta1} and \Ref{Theta3} we can thus try not only to
keep components with $j = \maxl'$ for $\maxl'<\maxl$, but also to
change the relative factors between the different components. In
general, the quadratic constraint \Ref{Theta3} will then fail to be
satisfied, unless the generators appearing in \Ref{gauging} form again
a closed algebra. However, this is relatively easy to ascertain by
direct inspection. We will make use of this trick in order to
establish the admissibility of the new gauge group $SO(8,\C)\subset
E_{8(8)}$ (with two inequivalent embeddings) for the maximal $N=16$
theory.

\section{Non-semisimple gaugings}

We first exemplify the boost method by deriving new non-semisimple gauge 
groups from the maximal $N=16$ theory with compact gauge group 
$G_0 = SO(8) \times SO(8)$. The first two of our examples are especially
important because they are directly related to the YM type maximal 
supergravities obtained by compactification of IIA and IIB supergravity 
on AdS$_3 \times S^7$.\footnote{For the type I theory, the
reduction has been performed explicitly in~\cite{CvLuPo00}. 
The KK spectra of the IIA/IIB theories on $S^7$ have been given
in~\cite{MorSam02}.} The YM gauge group is $SO(8)$ in both cases, and 
by the general result of \cite{NicSam03a} the corresponding CS gauge group 
must be the non-semisimple extension of $SO(8)$ by a 28-dimensional 
group of translations $T_{28}$ transforming in the adjoint of $SO(8)$; 
this is indeed one of the non-semisimple groups listed in \Ref{G1}.
Different non-semisimple, and in particular purely nilpotent, gaugings 
can be obtained by boosting $SO(8)\times SO(8)$ with other boost 
generators $N\in E_{8(8)}$. The different boostings correspond to 
different gradings, and the associated semisimple gauge subgroups 
can be read off from the $E_8$ Dynkin diagram, which we give below with 
our numbering of the simple roots. 


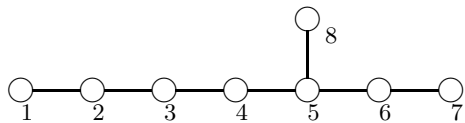
\setlength{\unitlength}{1973sp}%
\begingroup\makeatletter\ifx\SetFigFont\undefined
\def\x#1#2#3#4#5#6#7\relax{\def\x{#1#2#3#4#5#6}}%
\expandafter\x\fmtname xxxxxx\relax \def\y{splain}%
\ifx\x\y   
\gdef\SetFigFont#1#2#3{%
  \ifnum #1<17\tiny\else \ifnum #1<20\small\else
  \ifnum #1<24\normalsize\else \ifnum #1<29\large\else
  \ifnum #1<34\Large\else \ifnum #1<41\LARGE\else
     \huge\fi\fi\fi\fi\fi\fi
  \csname #3\endcsname}%
\else
\gdef\SetFigFont#1#2#3{\begingroup
  \count@#1\relax \ifnum 25<\count@\count@25\fi
  \def\x{\endgroup\@setsize\SetFigFont{#2pt}}%
  \expandafter\x
    \csname \romannumeral\the\count@ pt\expandafter\endcsname
    \csname @\romannumeral\the\count@ pt\endcsname
  \csname #3\endcsname}%
\fi
\fi\endgroup
\begin{center}
\begin{picture}(5716,1432)(1643,-1111)
{\thinlines
\put(7201,-736){\circle{300}}
}%
{\put(6301,-736){\circle{300}}
}%
{\put(5401,-736){\circle{300}}
}%
{\put(4501,-736){\circle{300}}
}%
{\put(3601,-736){\circle{300}}
}%
{\put(2701,-736){\circle{300}}
}%
{\put(1801,-736){\circle{300}}
}%
{\put(5401,164){\circle{300}}
}%
{\put(7051,-736){\line(-1, 0){600}}
}%
{\put(6151,-736){\line(-1, 0){600}}
}%
{\put(5251,-736){\line(-1, 0){600}}
}%
{\put(4351,-736){\line(-1, 0){600}}
}%
{\put(2551,-736){\line(-1, 0){600}}
}%
{\put(3451,-736){\line(-1, 0){600}}
}%
{\put(5401, 14){\line( 0,-1){600}}
}%
\put(1801,-1111){\makebox(0,0)[lb]{\smash{\SetFigFont{9}{10.8}{rm}{1}%
}}}
\put(2701,-1111){\makebox(0,0)[lb]{\smash{\SetFigFont{9}{10.8}{rm}{2}%
}}}
\put(3601,-1111){\makebox(0,0)[lb]{\smash{\SetFigFont{9}{10.8}{rm}{3}%
}}}
\put(4501,-1111){\makebox(0,0)[lb]{\smash{\SetFigFont{9}{10.8}{rm}{4}%
}}}
\put(5401,-1111){\makebox(0,0)[lb]{\smash{\SetFigFont{9}{10.8}{rm}{5}%
}}}
\put(6301,-1111){\makebox(0,0)[lb]{\smash{\SetFigFont{9}{10.8}{rm}{6}%
}}}
\put(7201,-1111){\makebox(0,0)[lb]{\smash{\SetFigFont{9}{10.8}{rm}{7}%
}}}
\put(5626,-136){\makebox(0,0)[lb]{\smash{\SetFigFont{9}{10.8}{rm}{8}%
}}}
\end{picture}
\end{center}
The most general grading is obtained by assigning real numbers $s_i$ 
(usually taken to be non-negative integers) to the simple roots $\Ga_i$ 
and defining the degree $\CD$ of a given root $\Ga=\sum_i n_i \Ga_i$ as
\be
\CD (\Ga) = \sum_j n_j s_j
\;.
\ee
The vector $(s_1,s_2, \dots)$ will be referred to as the 
``grading vector''. The roots of the semisimple subalgebra 
$\tilde \mathfrak{g}_0\subset \mathfrak{g}_0$ that is preserved 
by the boosting obviously satisfy $\CD (\Ga) =0$.

The extension of these considerations to lower $N$ is immediate.
In particular, the existence of gauged supergravities in four
dimensions with $n<8$ supersymmetries and gauge groups $SO(n)$
implies the existence of CS type gauged supergravities in three
dimensions with $N=2n$ supersymmetries and CS gauge groups
$SO(n) \ltimes T$ with ${\rm dim}\, T= \frac12 n(n-1)$. \footnote{The 
$n\leq 4$ and $n=5$ gauged theories in four dimensions were found already 
long ago in \cite{FreDas77,FreSch78} and \cite{deWNic81}, respectively. 
The $SO(6)$ gauged supergravity was never explicitly constructed, but 
can be obtained by truncation of the maximal $SO(8)$ gauged theory.}

\mathon
\subsection{$G_0 = SO(8)\ltimes T_{28}$ (type IIA)}
\mathoff

\label{IIAS7}

Type IIA supergravity can be compactified on AdS$_3 \times S^7$, 
giving rise to a maximal YM gauged supergravity with gauge group 
$SO(8)$, which furthermore coincides with the $S^1$ 
reduction of maximal $SO(8)$ gauged supergravity in four
dimensions. We now show how to obtain the required CS gauge group 
$G_0 = SO(8)\ltimes T_{28}$ from the compact gauge group 
$SO(8)\times SO(8)$ by an appropriate boost. In the next section, 
we will exhibit a second and inequivalent theory with the same 
gauge group based on the compactification of IIB supergravity
on AdS$_3 \times S^7$.

The construction is based on the 5-graded decomposition (i.e.\
$\ell=2$) of $E_{8(8)}$ 
under its subgroup $E_{7(7)} \times SL(2,\R)$
\be\label{IIA}
{\bf 248} = {\bf 1} \oplus {\bf 56} \oplus 
\big[{\bf 1} \oplus {\bf 133}\big] \oplus {\bf 56}\oplus {\bf 1} 
\;,
\ee
and associated with the grading vector
\be
(s_1,\dots , s_8)= (1,0,0,0,0,0,0,0) \;.
\ee

To give more details, we further decompose the generators w.r.t.\ the
embedding $\mathfrak{so}(8) \subset \mathfrak{sl}(8,\R)\subset 
\mathfrak{e}_{7(7)}$. 
Using $SO(8)$ indices $a, \Ga,\dot\Ga$ for the representations 
${\bf 8}_v$, ${\bf 8}_s$ and ${\bf 8}_c$, respectively, 
the $\mathfrak{e}_{8(8)}$ generators $X^{IJ}$ and $Y^A$ decompose as
\begin{eqnarray}
\label{IA2}
I &:& {\bf 8}_v + {\bf 8}_v \quad \Longrightarrow \quad
[IJ] \;\; : \;\; {\bf 28} + {\bf 28} + {\bf 1} +  {\bf 28} + {\bf 35}_v
\;,
\nonumber\\
A &:& {\bf 8}_s \times {\bf 8}_s + {\bf 8}_c \times {\bf 8}_c ~=~
{\bf 1} + {\bf 28}+ {\bf 35}_s +{\bf 1}+{\bf 28}+{\bf 35}_c \;.
\end{eqnarray}
Hence, $\mathfrak{e}_{8(8)}$ decomposes as (cf. \Ref{decA})
\be
{\bf 248} = {\bf 1} \oplus \big[{\bf 28} \oplus {\bf 28}\big] \oplus 
\big[ {\bf 1} \oplus {\bf 28} \oplus {\bf 35}_v \oplus {\bf 35}_s 
\oplus {\bf 35}_c \big] \oplus \big[{\bf 28} \oplus {\bf 28}\big] 
\oplus {\bf 1} 
\;.
\ee
Now we see that the representation content in \Ref{IIA} matches indeed 
with that of $N=8,D=4$ gauged supergravity \cite{dWiNic82} reduced 
on a torus: after removal of the $ {\bf 1} + 3\cdot {\bf 28} +
{\bf 35}_v$ gauge degrees of freedom, we are left with
the 70 scalars of the $D=4$ theory, and 2$\times$28 scalars coming 
from the 28 (YM) vector fields, together with two more scalar 
fields descending from the dilaton and the graviphoton, and
residing in the coset $SL(2,\R)/ SO(2)$.
The level zero generators consist of the grading generator $N$, 
the generators $E_{ab}= - E_{ba}$ and $S_{ab}=  S_{ba}$ which obey
\ba\label{comm1}
[E_{ab}, E_{cd}] &=& 2\delta_{d[a} E_{b]c}- 2\delta_{c[a} E_{b]d}
\;,\nonumber\\{} 
[E_{ab}, S_{cd}] &=& 2 \delta_{d[a} S_{b]c} + 
                   2\delta_{c[a}S_{b]d}\;,\nonumber\\{}
[S_{ab}, S_{cd}] &=& 2 \Gd_{d(a} E_{b)c} + 2 \Gd_{c(a} E_{b)d} 
\;,
\ea
and thus close into the Lie algebra of $SL(8,\R)\subset E_{7(7)}$,
together with the 35 compact and the 35 non-compact (traceless) 
generators $S_{\Ga\Gb}= S_{\Gb\Ga}$ and 
$S_{\dot\Ga\dot\Gb}= S_{\dot\Gb\dot\Ga}$, respectively, which enlarge 
$SL(8,\R)$ to the full $E_{7(7)}$. At level one, we have the 
28 + 28 generators $(U_{ab},V^{ab})$ transforming in the fundamental
${\bf 56}$ representation of $E_{7(7)}$, and at level $-1$ a conjugate
set of 28 + 28 generators $(U^{ab},V_{ab})$. These obey 
\be
[U_{ab}, U_{cd} ] = [V^{ab}, V^{cd} ] = 0 \;\; , \quad
[U_{ab} , V^{cd} ] = \Gd_{ab}^{cd}\, S^+ \;,
\ee
where $S^+$ is the level-2 singlet (the formulas for levels $-1$
and  $-2$ are analogous). The compact group $SO(8)\times SO(8)$ is
generated by the $E_{ab}$ building the diagonal subgroup, and the
linear combinations $U_{ab} - U^{ab}$ (or alternatively $V_{ab} -
V^{ab}$). Its embedding tensor has been given in
\cite{NicSam00,NicSam01a} and is of the type~\Ref{projector} with
the relative gauge coupling constant equal to $-1$. In the above
basis, it takes the form
\be 
\Theta = E_{ab}\vee U_{ab} - E_{ab} \vee U^{ab} \;,
\label{Theta88A}
\ee
(with the usual summation convention on the indices $a, b$)
as one can easily check by writing out \Ref{gauging} explicitly.
Therefore $\Theta$ has non-vanishing components only at grades
$\pm 1$.

Boosting with $N$, we get
\be\label{Thboost01}
\exp (\lambda N) \big(\Theta\big) =  
e^{\lambda} E_{ab} \vee U_{ab} - e^{-\lambda} E_{ab} \vee U^{ab} 
\;.
\ee
Rescaling and taking the limit $\Gl\rightarrow +\infty$ as described 
above, we find the new embedding tensor
\be\label{Thboost02}
\overline\Theta = E_{ab} \vee U_{ab}
\;.
\ee
Hence, the nonvanishing components of the new embedding tensor 
$\overline{\Theta}$ appear at grade  $+1$. The associated Lie
algebra is indeed the one corresponding 
to the group $SO(8)\ltimes T_{28}$, and is spanned by the 28 $SO(8)$ 
generators $E_{ab}$ and the 28 translation generators $U_{ab}$.
To see this even more explicitly, we write out the minimal
coupling~\Ref{gauging}
\be\label{ThB}
\Theta_{\cM\cN} B_\mu^{\;\cM} t^\cN = A_\mu^{ab} E_{ab} + C_\mu^{ab}
U_{ab} 
\;.
\ee
Here $A_\mu$ and $C_\mu$ are those 28+28 vector fields out of the 
248 vector fields ${B_\mu}^\cM$ that are `excited' by the gauging.

\mathon
\subsection{$G_0 = SO(8) \ltimes T_{28}$ (type IIB)}
\mathoff

\label{IIBS7}

Compactification of type IIB supergravity on AdS$_3 \times S^7$
gives 
rise to another maximal gauged supergravity of YM type in three dimensions 
with gauge group $SO(8)$.
This theory is again equivalent on shell to a 
maximal gauged supergravity of CS type with non-semisimple gauge group 
$SO(8)\ltimes T_{28}$. Although the gauge groups of the IIA and 
IIB compactifications are thus the same, their respective embeddings 
into $E_{8(8)}$ differ by a triality rotation, and there is no 
transformation that maps the two theories onto one another.
Unlike the IIA theory given above which has its alternative
origin in the maximal gauged D=4 theory of~\cite{dWiNic82},
the IIB theory may not
be obtained by simple torus reduction from higher dimensions.

The embedding of $SO(8)\times SO(8)$ into $E_{8(8)}$ in the IIB basis
and the identification of the requisite boost generator rely on the
7-graded decomposition (i.e. $\ell =3$) of $E_{8(8)}$ w.r.t.\ its 
$SL(8,\R)$ subgroup introduced in \cite{CJLP97} (see also appendix B
of \cite{KoNiSa00})
\be\label{IIB}
{\bf 248} = {\bf 8} \oplus \overline{\bf 28} \oplus {\bf 56} \oplus 
\big[{\bf 1} \oplus {\bf 63}\big]\oplus \overline{\bf 56}\oplus {\bf 28} 
\oplus \overline{\bf 8} \;.
\ee
The grade zero sector consists of the $\mathfrak{sl}(8,\R)$ subalgebra
and a singlet which will serve as the boost (and grading) generator.
The grading vector is 
\be
(s_1, \dots , s_8) =  (0,0,0,0,0,0,0,1)\;.
\ee 
Next we decompose the generators w.r.t.\ the subgroup
$SO(8)\subset SL(8,\R)$ which gives (cf. \Ref{decB})
\be\label{decompSO8}
{\bf 248} = {\bf 8}_v \oplus {\bf 28} \oplus {\bf 56}_v \oplus 
\big[{\bf 1} \oplus {\bf 28} \oplus {\bf 35}_v \big]\oplus {\bf 56}_v
\oplus {\bf 28} \oplus {\bf 8}_v \;.
\ee
Indeed, this matches with the lowest floor of the KK tower of the IIB
theory on $S^7$~\cite{MorSam02}. Using the same notations as in the
foregoing section, the $E_{8(8)}$ generators $X^{IJ}$ and $Y^A$ now
decompose as (cf. \Ref{decA})
\begin{eqnarray}\label{IA1}
I &:& {\bf 8}_s + {\bf 8}_c \quad \Longrightarrow \quad
[IJ] \;\; : \;\; {\bf 28} + {\bf 28} + {\bf 8}_v + {\bf 56}_v
\;,
\nonumber\\
A &:& {\bf 8}_v \times {\bf 8}_v + {\bf 8}_s \times {\bf 8}_c ~=~
{\bf 1} + {\bf 28}+ {\bf 35}_v +{\bf 8}_v + {\bf 56}_v
\;.
\end{eqnarray}
The diagonal $SO(8)$ subgroup is generated by the elements (see
appendix B of \cite{KoNiSa00} for notations)
\be\label{E}
E_{ab} :=
\ft14 \left( \gamma^{ab}_{\alpha\beta}\,X^{\alpha\beta} +
\gamma^{ab}_{\dot{\alpha}\dot{\beta}}\,X^{\dot{\alpha}\dot{\beta}}\right)
\;,
\ee
with $SO(8)$ $\gamma$-matrices, and where $X^{\alpha\beta}$ and
$X^{\dot{\alpha}\dot{\beta}}$ generate 
the compact $SO(8) \times SO(8)$ subgroup of $E_{8(8)}$. The commutation
relations are the same as in \Ref{comm1}.

The grading operator $N\equiv Y^{cc}$ commutes with $\mathfrak{sl}(8,\R)$ 
(so we have in particular $[N, E_{ab} ] = 0 $), and therefore this 
subalgebra is unaffected by any boosting with $N$. We also need the 
28+28 nilpotent abelian generators (cf. appendix B of \cite{KoNiSa00})
\ba\label{Z}
Z_{ab} &=& \ft18 \left(\gamma^{ab}_{\alpha\beta}\,X^{\alpha\beta} -
\gamma^{ab}_{\dot{\alpha}\dot{\beta}}\,X^{\dot{\alpha}\dot{\beta}} \right)
  + Y^{[ab]} \;,\nonumber\\
Z^{ab} &=& - \ft18 \left(\gamma^{ab}_{\alpha\beta}\,X^{\alpha\beta} -
\gamma^{ab}_{\dot{\alpha}\dot{\beta}}\,X^{\dot{\alpha}\dot{\beta}} \right)
  + Y^{[ab]}\;.
\ea
The relevant commutation relations between these generators are
\ba
[E_{ab}, E_{cd}] &=& 2\delta_{d[a} E_{b]c}- 2\delta_{c[a}
E_{b]d}\;,\nonumber\\{} 
[E_{ab}, Z_{cd}] &=& 2 \Gd_{d[a} Z_{b]c} - 2 \Gd_{c[a}
Z_{b]d}\;,\nonumber\\{}
[E_{ab}, Z^{cd}] &=& 2 \Gd^{d[a} Z^{b]c} - 2 \Gd^{c[a}
Z^{b]d}\;,\nonumber\\{}
[Z_{ab}, Z_{cd}] &=& [Z^{ab}, Z^{cd}] = 0 \;,
\label{commB}
\ea
together with
\be\label{NZ}
[N, Z_{ab}] = 2 Z_{ab} \;\; , \qquad
[N, Z^{ab}] = - 2 Z^{ab}\;.
\ee

In this basis, the compact gauge group $SO(8)\times SO(8)$ is
generated by the $E_{ab}$ building the diagonal subgroup, and
the linear combinations $Z_{ab}-Z^{ab}$. The embedding
tensor~\Ref{Theta88A} in this basis is
\be\label{ThetaEZ}
\Theta = E_{ab} \vee Z_{ab} - E_{ab} \vee Z^{ab} \;,
\ee
with all other components of $\Theta$ vanishing. Therefore, the 
embedding tensor has nonvanishing components only at levels $\pm 2$.

As before, we boost with $N$ and make use of \Ref{NZ}, to get
the new embedding tensor
\be\label{Thboost2}
\overline\Theta = E_{ab} \vee Z_{ab} \;.
\ee
In accordance with out general arguments above $\overline\Theta$ thus has 
only the graded piece $\overline\Theta{}^{(2)}$. That
\Ref{Thboost2} is 
indeed the embedding tensor for $SO(8)\ltimes T_{28}$ inside 
$E_{8(8)}$ follows as in \Ref{ThB}.

\mathon
\subsection{$G_0 = SO(p,8-p)\ltimes T_{28}$}
\mathoff

$N=8$ supergravity in four dimensions admits gaugings for all gauge 
groups $CSO(p,q;r)$ with $p+q+r=8$ \cite{Hull84a,Hull84b,Hull84c}. 
These are semisimple for $r=0$, and non-semisimple for $r>0$ (and 
all contained in $E_{7(7)}$ regardless of the choice of $p,q,r$). 
By dimensional reduction on $S^1$, each of these theories gives 
rise to a maximal gauged supergravity of YM type in three dimensions, 
whose CS type description requires the gauge groups 
$G_0 = CSO(p,q;r)\ltimes T \subset E_{8(8)}$ such that the elimination 
of the translational gauge fields associated with $T$ leads back 
to a YM type gauged supergravity with gauge group $CSO(p,q;r)$. 
We first discuss the case $r=0$, which descends from the semisimple 
non-compact gaugings with gauge groups $SO(p,8-p)$ in four dimensions. 

As explained in the introduction, there is no distinction between
IIA and IIB for these non-compact embeddings. For definiteness,
we will therefore use the IIB basis of section~\ref{IIBS7}, where we
already defined the $SO(8)$ generators $E_{ab}$ and the nilpotent 
generators $Z_{ab}$ and $Z^{ab}$. We also need the 35 non-compact generators
\be\label{S}
S_{ab} := 2 Y^{(ab)} - \ft14 \Gd^{ab} Y^{cc}\;,
\ee
which enlarge $\mathfrak{so}(8)$ to $\mathfrak{sl}(8,\R)$. In addition
to \Ref{commB} we have the commutation relations
\ba
[S_{ab}, Z_{cd}] &=& -2 \Gd_{d(a} Z_{b)c} + 2 \Gd_{c(a} Z_{b)d}
                     - \ft12 \Gd_{ab} Z_{cd} \;,\nonumber\\{} 
[S_{ab}, Z^{cd}] &=& - 2 \Gd^{d(a} Z^{b)c} + 2 \Gd^{c(a} Z^{b)d}
                     - \ft12 \Gd_{ab} Z^{cd}
                     \;,
\ea
and
\be
[Z_{ab} , Z^{cd}] = \Gd_{c[a} \big( E_{b]d} + S_{b]d}) 
    - \Gd_{d[a} \big( E_{b]d} + S_{b]c}) + \ft12 \Gd_{ab}^{cd}
N \;.
\ee
To identify the $SO(p,q)\times SO(p,q)$ subgroup inside $E_{8(8)}$, 
we split the $SO(8)$ indices $a,b,...$ into indices 
$i,j,... \in \{1,\dots, p \}$ and $r,s,... \in \{p+1, \dots,8\}$. 
Then the Lie algebras of the two factor groups $SO(p,q)$ are spanned by
\ba
&& E_{ij} - Z_{ij} + Z^{ij} \;\; , \quad E_{rs} + Z_{rs} - Z^{rs}
   \;\; , \quad S_{ir} + Z_{ir} + Z^{ir} \;,\nonumber\\
&& E_{ij} + Z_{ij} - Z^{ij} \;\; , \quad E_{rs} - Z_{rs} + Z^{rs}
   \;\; , \quad S_{ir} - Z_{ir} - Z^{ir}\;,
\ea
respectively, where the generators $E_{ij} \pm (Z_{ij} - Z^{ij})$ and
$E_{rs} \pm (Z_{rs} - Z^{rs})$ are compact, whereas the $pq$ generators
$S_{ir} \pm (Z_{ir} + Z^{ir})$ are non-compact. 
In this basis, the embedding tensor of the maximal gauged
supergravity with gauge group $SO(p,q) \times 
SO(p,q)$~\cite{NicSam01a} is given by
\be\label{SOpq}
\Theta = E_{ij}\vee \big( Z_{ij} - Z^{ij} \big) - 
         E_{rs}\vee \big( Z_{rs} - Z^{rs} \big) +
         S_{ir}\vee \big( Z_{ir} +  Z^{ir} \big)\;.
\ee
Applying the boost method as before we get the new embedding tensor
\be
\overline\Theta= E_{ij} \vee Z_{ij} - E_{rs} \vee Z_{rs} +
                 S_{ir}\vee Z_{ir} \;,
\ee
corresponding to the non-semisimple group $G_0 = SO(p,q) \ltimes T_{28}$
whose $SO(p,q)$ subgroup is generated by $\{ E_{ij}, E_{rs}, S_{ir}\}$,
and whose nilpotent part is spanned by the 28 elements $Z_{ab}$. 

In \cite{FiNiSa02} we have given the embedding tensor~\Ref{SOpq}
in the IIA basis in terms of $SO(8)$ $\gamma$-matrices. The
above IIB basis is triality rotated w.r.t.\ to the one used there,
which explains the simpler form of the above embedding tensor.

\mathon
\subsection{$G_0 = CSO(p,q;r)\ltimes T_{p,q,r}$ for $r>0$}
\mathoff

For $r>0$ the non-semisimple groups $CSO(p,q;r) \times CSO(p,q;r)$ 
cannot be embedded into  $E_{8(8)}$ and thus are {\em not} 
admissible gauge groups for $N=16$ supergravity. However, there  
are the non-semisimple groups containing only one $CSO(p,q;r)$ factor,
namely the groups $CSO(p,q;r)\ltimes T_{p,q,r}$ from the list \Ref{G1}.
As there seems to be no way to get these non-semisimple gauge groups 
by the boost method, we proceed directly to their description.
For this purpose, we have to further refine the split of $SO(8)$ 
indices $a,b,...$ into $i,j,... \in \{1,\dots, p\}$, $m,n,...\in 
\{p+1,\dots , p+q\}$ and $s,t,... \in \{p+q+1, \dots,8\}$. The 
generators of the non-semisimple subgroup $CSO(p,q;r)$ are then 
$E_{ij}$, $E_{mn}$ and $S_{im}$ for the non-compact semisimple 
subgroup $SO(p,q)$, and the $(p+q)r$ translation generators
\ba
T_{is} &:=& E_{is} + S_{is} \;,\qquad
T_{ms} ~:=~ E_{ms} + S_{ms} \;,
\ea
both of which can be obtained by boosting $E_{is}$ and $E_{ms}$
within $E_{7(7)}$. To this algebra we adjoin the $\ft12 (p+q)(p+q-1) + (p+q)r$
commuting generators $Z_{ij}, Z_{mn}, Z_{im}$ and $Z_{is}$,
$Z_{ms}$:
this defines the Lie algebra of the non-semisimple group 
$CSO(p,q;r)\ltimes T_{p,q,r}$. The removal of the generators
$E_{st}$ 
from the definition of $CSO(p,q;r)$ is thus accompanied by the removal 
of the nilpotent elements $Z_{st}$ from the translation group $T_{28}$
yielding the subgroup $T_{p,q,r}$. Together with the  $(p+q)r$ nilpotent 
generators of $CSO(p,q;r)$, we thus have altogether 
$\ft12(p+q)(p+q-1) + 2(p+q)r$ nilpotent generators. Although this
number may exceed 36 (for instance with the choice $(p,q,r)=(3,2,3)$), 
which is the maximal number of mutually commuting 
nilpotent generators in $E_{8(8)}$, there is no contradiction because
\ba
[ T_{is} , Z_{jt} ] &=& \delta_{st} Z_{ij} \;,\qquad
[ T_{ms} , Z_{nt} ] ~=~ \delta_{st} Z_{mn} \;,
\ea
do not commute. Observe again that $Z_{st}$ does not appear in this 
commutation relation. The maximal number of mutually commuting generators 
in  $CSO(p,q;r)\ltimes T_{p,q,r}$ is thus equal to 
$\ft12 (p+q)(p+q-1) + (p+q)r < 28$ for any choice of $(p,q,r)$.

Analysis of the projector condition \Ref{27000} reveals that the
$CSO(p,q;r)\ltimes T_{p,q,r}$ embedding tensor has the following 
non-vanishing components 
\ba
\overline\Theta &=& E_{ij} \vee Z_{ij} - E_{mn} \vee Z_{mn} +
   2 S_{im} \vee Z_{im} \nonumber\\
    && + T_{is} \vee Z_{is} -  T_{ms} \vee Z_{ms}
    \;,
\ea
with all other components vanishing.

\mathon
\subsection{Nilpotent gauge groups}
\mathoff

There are many other gradings which one can use to boost
$SO(8)\times SO(8)$ or the non-compact and non-semisimple
gauge groups discussed in the foregoing sections. However, most 
of these will lead to nilpotent gauge groups because the compact 
part of the gauge group is boosted away completely. Moreover,
different boosts do not necessarily lead to new and different 
gauge groups. For instance, due to the high symmetry of the 
$SO(8)\times SO(8)$ embedding tensor, the more difficult task is 
to obtain {\em different} gaugings from boosting from that group.

To give an example, still with maximal grade zero subalgebra, choose
the grading vector $(s_1,\dots , s_8)= (0,0,0,0,0,0,1,0)$,
which yields the 5-graded decomposition
\be
{\bf 248} = {\bf 14} \oplus {\bf 64} \oplus 
\big[ {\bf 1} \oplus {\bf 91} \big]\oplus {\bf 64}\oplus {\bf
14} 
\;.
\ee
Inspection of the Dynkin diagram shows that the $\bf 91$ at grade
zero is the $\mathfrak{so}(7,7)$ algebra, under which the $\bf 14$
and $\bf 64$ transform as the vector and spinor representations,
respectively. With regard to this subalgebra, the embedding tensor 
has non-vanishing graded pieces 
\be
\GTh \in \mathfrak{g}^{(2)} \vee \mathfrak{g}^{(2)} \oplus
         \mathfrak{g}^{(2)} \vee \mathfrak{g}^{(-2)}\oplus
         \mathfrak{g}^{(0)} \vee \mathfrak{g}^{(0)}\oplus
         \mathfrak{g}^{(-2)} \vee \mathfrak{g}^{(-2)}\;.
\ee
After boosting, we are thus left with a purely nilpotent embedding tensor
\be 
\overline\GTh \in \mathfrak{g}^{(2)} \vee \mathfrak{g}^{(2)} \;,
\ee
yielding a 14-dimensional abelian nilpotent gauge group. We can
identify this group with the gauge group 
\be
G_0 = CSO(1,0;7) \times T_{1,0,7}
\;,
\ee
of the previous section.

Choosing instead the grading vector $(s_1,\dots , s_8)= (1,0,0,0,0,0,1,0)$
gives rise to the 9-graded decomposition 
\bea
{\bf 248} &=& {\bf 1} \oplus {\bf 12} \oplus \big[{\bf 32} \oplus {\bf 1}\big] 
\oplus \big[ {\bf 32} \oplus {\bf 12} \big]  \oplus 
\big[ {\bf 1} \oplus {\bf 1}\oplus {\bf 66}\big] \nonumber\\
&& \oplus \big[ {\bf 32} \oplus {\bf 12}\big] \oplus 
\big[{\bf 32} \oplus {\bf 1}\big] \oplus {\bf 12} \oplus {\bf
1}
\;,
\eea
with an $\mathfrak{so}(6,6)$ and {\it two} singlets in the middle, and
the vector representation $\bf{12}$ and the spinor representation $\bf{32}$.
There are thus two possible boost generators, whose different linear 
combinations correspond to the different values of $s_1$ and $s_7$ 
(both of which have been chosen $=1$ above). Now, $\Theta$ decomposes as
\ba
\GTh &\in& \mathfrak{g}^{(4)} \vee \mathfrak{g}^{(2)} \, \oplus \, 
           \mathfrak{g}^{(3)} \vee \mathfrak{g}^{(3)} \,\oplus\,
           \mathfrak{g}^{(4)} \vee \mathfrak{g}^{(-2)}\, \oplus \,
           \mathfrak{g}^{(1)} \vee \mathfrak{g}^{(1)}\nonumber\\
      &&   \oplus \; \mathfrak{g}^{(3)} \vee \mathfrak{g}^{(-3)}\, \oplus \,
           \mathfrak{g}^{(0)} \vee \mathfrak{g}^{(0)}\, \oplus\,
           \mathfrak{g}^{(-1)} \vee \mathfrak{g}^{(-1)} \nonumber\\
    &&   \oplus \;  \mathfrak{g}^{(2)} \vee \mathfrak{g}^{(-4)}\, \oplus\,
          \mathfrak{g}^{(-3)} \vee \mathfrak{g}^{(-3)} \, \oplus \,
           \mathfrak{g}^{(-4)} \vee \mathfrak{g}^{(-2)} \;.
\ea
Boosting with the above grading vector leaves us with
\be
\overline\GTh \in \mathfrak{g}^{(4)} \vee \mathfrak{g}^{(2)} \oplus
                  \mathfrak{g}^{(3)} \vee \mathfrak{g}^{(3)}\;.
\ee
Examining the representation content of the grade 6 contributions to
$\overline\GTh$, one sees that the associated nilpotent gauge group
contains only the 14 = 1+12+1 nilpotent generators, and therefore
coincides with the gauge group $CSO(1,0;7)\times T_{1,0,7}$ obtained 
above.

\mathon
\section{Complex gaugings: $G_0 = SO(8,\C)$}
\mathoff

We now come to our most surprising result, which is the admissibility
of the {\em complex} group $G_0 = SO(8,\C)$. This result is arrived at 
by exploiting the observation made at the end of section~2, according 
to which we can change the relative factors between the components of 
the embedding tensor at different grades, as long as the modified 
embedding tensor still defines a closed algebra. Here we simply 
need to switch the relative sign between the two terms in the 
$SO(8) \times SO(8)$ embedding tensor in \Ref{ThetaEZ} to get
\be\label{ThetaSOC}
\Theta = E_{ab} \vee Z_{ab} + E_{ab} \vee Z^{ab}
\;,
\ee
again with all other components vanishing. Writing out \Ref{gauging}
with \Ref{ThetaSOC} one immediately deduces that the associated 
Lie algebra is spanned by the $\mathfrak{e}_{8(8)}$ elements $E_{ab}$ 
and the 28 elements
\be
F_{ab} := Z_{ab} + Z^{ab} \equiv 2 Y^{[ab]}
\;,
\ee
which together again form a closed algebra:
\ba
[ E_{ab}, E_{cd} ] &=&  
     2\delta_{d[a} E_{b]c}- 2\delta_{c[a} E_{b]d}\;,\nonumber\\{}
[ E_{ab}, F_{cd} ] &=&  
     2\delta_{d[a} F_{b]c}- 2\delta_{c[a} F_{b]d}\;,\nonumber\\{}
[ F_{ab}, F_{cd} ] &=&  
    - 2\delta_{d[a} E_{b]c} + 2\delta_{c[a} E_{b]d}\;.
\ea
Note the relative minus sign between the first and the third line.
It is easy to see that this Lie algebra is isomorphic to the complex 
Lie algebra $\mathfrak{so}(8,\C)$, if we we decree the generators 
$E_{ab}$ to be `real' and the generators $F_{ab}$ to be `imaginary'
(so the latter can be thought of as `$iE_{ab}$'). A second 
$\mathfrak{so}(8,\C)$ is obtained by replacing the IIB generators 
$Z_{ab}$ and $Z^{ab}$ by the corresponding IIA generators $U_{ab}$ 
and $U^{ab}$.

Under the action of $\mathfrak{so}(8,\C)$ in the IIB basis the adjoint
${\bf 248}$ of $E_{8(8)}$ decomposes into three irreducible subspaces:
the first of these is the $\mathfrak{so}(8,\C)$ subalgebra itself, the
second is the 64-dimensional subspace spanned by the generators $
Z_{ab} - Z^{ab}$, $S_{ab}$ (cf. \Ref{S}) and the grading operator $N$,
and the third is the 128-dimensional subspace spanned by the level
$\pm 3$ and $\pm 1$ generators in \Ref{IIB} (i.e.  the generators
$Z^a, Z_a, E^{abc}, E_{abc}$ in the notation of
\cite{CJLP97,KoNiSa00}). For the IIA basis we find that the ${\bf
248}$ decomposes instead into the subalgebra and {\em three}
64-dimensional irreducible subspaces. To understand these rather
counterintuitive results, we recall that there exist simple real forms
of Lie algebras whose complexification is no longer simple.\footnote{A
familiar example is the Lorentz group, where
$$
\C\otimes\mathfrak{so}(1,3) = \mathfrak{so}(4,\C) = 
\mathfrak{so}(3,\C)\oplus \mathfrak{so}(3,\C)\;.
$$} In the case at hand, $\mathfrak{so}(8,\C)$ is embedded as a simple
Lie algebra into $\mathfrak{e}_{8(8)}$, but its complexification
in  $\mathfrak{e}_{8}(\C)$ is no longer simple:  
\be
\C\otimes\mathfrak{so}(8,\C) =
\mathfrak{so}(8,\C)'\oplus \mathfrak{so}(8,\C)' \;,
\ee
where the prime on the r.h.s.\ is to indicate two copies of the standard
complexified $\mathfrak{so}(8)$. The 64-dimensional irreducible subspace 
just identified may then be viewed as a real section of the complex 
$(\bf{8},\bf{8})$ representation of $SO(8,\C)\times SO(8,\C)$. 

In terms of the $SO(16)$ decomposition \Ref{Theta} the $SO(8,\C)$ 
embedding tensor is purely `off-diagonal', viz.
\be
\Theta_{IJ|A} = -\ft17 \Gamma^{[I}_{A \dot A} \Xi^{J]\dot A} \;\; , 
\qquad \GTh_{IJ|KL} = \GTh_{A|B} =0 \;,
\ee
where, in the $SO(8)$ decomposition \Ref{IA1}, $\Xi^{I\dot A}$ has the 
non-vanishing components 
\be
\Xi^{I \dot A} =  \left\{ \begin{array}{rl}
\gamma^a_{\Ga\dot\Ga} & \mbox{for $I=\Ga$ and $\dot A = (a\dot\Ga)$} \\
- \gamma^a_{\Ga\dot\Ga} & \mbox{for $I=\dot\Ga$ and $\dot A = (a\Ga)$}
                      \end{array}
                      \right.
                      \;.
\ee
The relative sign is fixed by requiring $\Gamma^I_{A\dot A}\Xi^{I \dot A}=0$.
Consequently the vacuum expectation values of both $A_1^{IJ}$ 
and $A_3^{\dot A \dot B}$ vanish at the origin, see \Ref{AinT};
from \Ref{potential} we immediately obtain
\be
\langle W \rangle = + \ft1{16} g^2 A_2^{I\dot A}  A_2^{I\dot A} 
    = \ft1{16} g^2 \Xi^{I\dot A}  \Xi^{I\dot A} = 8 g^2  > 0
\;,
\ee
which is a de Sitter vacuum with completely broken supersymmetry,
where the $SO(8,\C)$ symmetry is broken to its compact subgroup 
$SO(8)\equiv SO(8,\R)$. The fermionic mass term is purely off-diagonal
\be
\CL_m^{(f)} = \ft17 ig \Xi^{I\dot A} 
   \overline\chi^{\dot A} \gamma^\mu \psi_\mu^I
\;.
\ee
An analysis of the scalar mass matrix \cite{FiNiSa02}
\be
\CM^2_{AB} = -\ft3{16} g^2 \left( \Gamma^I_{A \dot A} \Xi^{J\dot A}
           \Xi^{J\dot B} \Gamma^I_{\dot B B} - 
 \Gamma^I_{A \dot A} \Xi^{J\dot A} \Xi^{I\dot B} \Gamma^J_{\dot B B}\right)
\;,
\ee
yields the (mass)$^2$ eigenvalues
\ba\label{massIIA}
\begin{tabular}{l||c|c|c}
$m_S^2$ & $16g^2$ & $0$ & $-48 g^2$ \\ \hline
$SO(8)$ & ${\bf 35}_s+{\bf 35}_c$ & ${\bf 28}+{\bf 28}$ & ${\bf
1}+{\bf 1} $
\end{tabular}
\;,
\ea
and
\ba\label{massIIB}
\begin{tabular}{l||c|c|c|c|c}
$m_S^2$ & $16g^2$ & $12g^2$ & $0$ & $-20g^2$ & $-48 g^2$ \\ \hline
$SO(8)$ & ${\bf 35}_v$ & ${\bf 56}_v$ & ${\bf 28}$ &${\bf 8}_v$ &
${\bf 1} $
\end{tabular}
\;,
\ea
for the IIA and the IIB embedding, respectively, in accordance
with the spectrum of representations \Ref{IA2}, \Ref{IA1}. The 
fact that these spectra come out to be different confirms the 
inequivalence of the IIA and IIB embeddings of $SO(8,\C)$ 
into $E_{8(8)}$. Because of the tachyonic directions present for
both embeddings, the de Sitter vacua are unstable. Moreover, a 
preliminary analysis indicates that neither potential has any 
non-trivial stationary points: in fact, numerical checks suggest
that the potential is a monotonic function along any geodesic
starting from the origin $\CV={\bf 1}$ in the scalar field space.

Starting from the $SO(p,q) \times SO(p,q)$ generators in either the 
IIA or the IIB basis, one finds the following alternative bases for 
$\mathfrak{so}(8,\C)$ in $E_{8(8)}$:
\ba\label{SOpqC}
E_{ij}, E_{rs}, S_{ir} \qquad && ({\rm real} \;{\rm generators}) \;,
\nonumber\\
Z_{ij} + Z^{ij}, Z_{rs} + Z^{rs}, Z_{ir} - Z^{ir} 
   \quad&& ({\rm imaginary} \; {\rm generators})
\;,
\ea 
with the embedding tensor
\be\label{SOpqC1}
\Theta = E_{ij}\vee \big( Z_{ij} + Z^{ij}\big)
       - E_{rs}\vee \big( Z_{rs} + Z^{rs}\big)
       + S_{ir}\vee \big( Z_{ir} - Z^{ir}\big)
\;.
\ee
It might thus appear that there are more inequivalent `$SO(p,q,\C)$ 
gaugings', but this is not the case -- in agreement with the well known 
fact that there is only one complex group over $SO(8)$ (which might 
be embedded in inequivalent ways, though, as we have seen). However, an 
analysis of the scalar mass spectrum at the origin reveals that the 
putative $SO(p,q,\C)$ theories have identical spectra: in the IIB basis 
adopted in \Ref{SOpqC1}, the scalar mass spectrum coincides with 
\Ref{massIIB} for even $p,q$, and with \Ref{massIIA} for odd $p,q$,
and vice versa for the type IIA gauging of $SO(8,\C)$. There should
thus exist an explicit $E_{8(8)}$ transformation relating the different 
but equivalent bases.

Are there similar complex gaugings for other values and dimensions? 
Let us first note that complex embeddings analogous to the one discussed 
above (i.e. without an imaginary unit $i$) exist for all split real forms. 
More specifically, for $\C^\times = \C, \C'$ or $\C^0$, and in analogy 
with the embedding $SO(8,\C^\times) \subset E_{8(8)}$, we have
\ba
SO(6,\C^\times) &\subset& E_{7(7)} \;, \nonumber\\
SO(5,\C^\times) &\subset& E_{6(6)} \;, \nonumber\\
SO(4,\C^\times) &\subset& E_{5(5)}\equiv SO(5,5) \;,
\ea
as one can show by truncating the decomposition \Ref{IIB} to the
relevant $SO(n)$ subgroups. These are the noncompact real forms
appearing in $D\geq 4$ supergravities \cite{CreJul79}, but a quick
counting argument shows that none of the groups on the l.h.s. are
viable gauge groups for these theories (for instance,
$SO(6,\C^\times)$ would require 30 vector fields in four dimensions). 
By contrast, in three dimensions we expect complex gaugings to 
exist for lower $N=2n$, because the existence of non-semisimple
gaugings with groups $SO(n)\ltimes T$ can be inferred from the
existence of corresponding gauged theories in four dimensions. 
For instance, $SO(6)\ltimes T_{15}$ may be embedded in the isometry
group of the coset space $E_{7(-5)}/(SO(12)\times SU(2))$ of the
three-dimensional $N=12$ theory; however, this embedding will
necessarily break the compact $SU(2)$ factor, in accordance
with the fact that the gauged theory has no global $SU(2)$ symmetry.
Flipping signs as in \Ref{ThetaSOC} will then produce the desired
gauged theories with $SO(n,\C)$.

\section{Potentials and supersymmetry breaking}

It has been known for some time that dimensionally reduced gauged
supergravities in general do not admit maximally supersymmetric ground
states, even if the ancestor theories do possess such vacua. A prime
example is the maximal $N=8$ theory in four dimensions \cite{dWiNic82}
which after torus reduction to three dimensions only admits a
partially supersymmetric domain wall solution~\cite{LuPoTo97}. In this
section, we would like to explain how this `loss of supersymmetric
vacuum' comes about by studying how the scalar potentials are affected
when a semisimple gauge group is replaced by a non-semisimple one.

As explained in section~2.2, the scalar potential $W$ \Ref{potential} 
is expressed in terms of the $T$-tensor~\Ref{Ttensor}. Like $\Theta$, 
the latter decomposes into a sum of graded pieces
\be
T_{\cA\cB} = \sum_{j=-2\ell}^{j=2\ell}T_{\cA\cB}^{(j)} \;\; , \qquad
T_{\cA\cB}^{(j)} :=
\CV^\cM_{\;\; \cA} \CV^\cN_{\;\;\cB} \Theta_{\cM\cN}^{(j)}
\;.
\la{T7}
\ee
The potential itself is a quadratic function of the $T$-tensor, and
can therefore be decomposed into graded pieces with grades ranging 
between $-4\ell$ and $+4\ell$ (cf.~\Ref{gradeTheta})
\be\label{potential1}
W = \sum_{n=-4\ell}^{4\ell} W^{(n)} \;,
\ee
where each term $W^{(n)}$ receives contributions from the products 
$T^{(j)} T^{(k)}$ with $j+k =n$. Consequently, under a rescaling 
with the boost generator $N$, we obtain
\be
W \longrightarrow \sum_{n=-4\ell}^{4\ell} e^{n\lambda} W^{(n)}
\;.
\ee
For those non-semisimple gaugings which can be generated
by the boost method, or by restricting the embedding tensor
to a given grade, the corresponding potentials can be immediately
deduced by replacing $\Theta$ by $\overline\Theta$
\be\label{T2}
\overline{T}_{\cA\cB} = \CV^\cM_{\;\; \cA} \CV^\cN_{\;\;\cB} 
\overline\Theta_{\cM\cN}
\;.
\ee
The singular boost leading to the non-semisimple gauge group thus
results in the removal of certain terms from the $T$-tensor, and
therefore in the removal of certain terms from the potential itself: 
after rescaling and taking the limit, we are left with a truncated 
potential 
\be\label{potential3}
\overline{W} = W^{(2\maxl)} \;.
\ee
computed from \Ref{T2} with the maximal grade $\maxl$ from
\Ref{Thetaboost2}.  It is this removal of lower grade contributions
which may `destabilize' a potential which originally did possess a
stable groundstate. Roughly speaking, the removal of certain terms
from the potential turns an initially `cosh-like' potential into an
exponential one, thus inducing a run-away behavior in special
directions in the scalar field manifold.

In order to further analyze this decomposition of the potential and to
elucidate the relation between the potentials obtained directly in
three dimensions and those obtained by dimensional reduction from
higher dimensional gauged supergravities, we define the `dilaton' to
be the scalar field $\phi$ associated with the grading generator $N$
by extracting its dependence from the 248-bein
\be\label{tildeV}
\CV(\tilde\phi, \phi) = \tilde\CV(\tilde\phi)\cdot \exp (\phi N)
\;.
\ee
This  decomposition requires that we choose a basis $\{ t^\cM \}$ 
of $\mathfrak{g}$ which is compatible with (i.e. diagonal w.r.t.)
the grading~\Ref{grading} with grades $d_\cM \equiv \CD (t^\cM)$, 
such that
\ba
t^\cM \in \mathfrak{g}^{(d_\cM)} \;.
\ea
The coset space ${\rm G}/{\rm H}$ may then be parametrized in a
triangular gauge by exponentiating the nilpotent positive-grade
generators $\{ t^{\cM} \,|\, d_\cM\!>\!0 \}$ together with the
non-compact generators at grade $d_\cM =0$. Corresponding to the grade
of their generators, we may then assign a charge to the scalar
fields. In the representation \Ref{tildeV}, the matrix $\tilde\CV$ is
an exponential containing only non-negative grade generators other
than $N$ with their associated fields $\tilde\phi$ (for which we will
not need an explicit parametrization). We shall verify below that for
those theories descending from higher dimensions, the field $\phi$ can
indeed be identified with the usual dilaton which is defined as the
ratio of metric determinants
\be 
\sqrt{g_{D}} = \sqrt{g_{3}}\, e^\phi \;,
\label{dilaton}
\ee
where $g_D$ and $g_3$ are the metric determinants in $D$ and three
dimensions, respectively. The parametrization~\Ref{tildeV}
correspondingly yields a `free' kinetic term $\propto 
\partial_\mu\phi\partial^\mu\phi$ for the dilaton $\phi$, whereas the
kinetic terms for the other fields $\tilde{\phi}$ come with a field
dependent metric and certain powers of $e^\phi$ depending on the
respective charges of the~$\tilde{\phi}$.

Defining the dilaton independent part of the $T$-tensor
\be\label{Ttilde} \tilde T_{\cA\cB} (\tilde\phi)= \tilde\CV^\cM_{\;\;
\cA} \tilde\CV^\cN_{\;\;\cB} \Theta_{\cM\cN} \;, \ee the dilaton
dependence of the potential can be made completely explicit. To this
aim, we first note that one must be careful in distinguishing for
every expression between its `dilaton power' (i.e.\ the integer $n$
appearing in the factor $e^{n\phi}$ multiplying this expression) and
its grade w.r.t.\ $N$: they are not the same, because, in \Ref{tildeV}
the grading operator $N$ acts {\em from the left} on $\CV$ whereas the
dilaton is factored out {\em on the right} of $\CV$. Hence, the matrix
$\CV^{\cM}{}_{\cA}$ decomposes as
\ba
\CV^{\cM}{}_{\cA} &=& \tr[\CV^{-1}
t^{\cM} \CV\, t_{\cA}] ~=~ e^{-d_\cA
\phi}\,\tilde{\CV}^{\cM}{}_{\cA}(\tilde{\phi})
\;.
\ea
Here $\tilde{\CV}^{\cM}{}_{\cA}$ no longer depends on $\phi$, its
grade w.r.t.\ $N$ is $d_\cM$, and it has charge $(d_\cA\!-\!d_\cM)$;
in particular, $\tilde{\CV}^{\cM}{}_{\cA}=0$ for $d_\cA<d_\cM$ by
triangularity. Similarly, the expansion~\Ref{T7} of the $T$-tensor
takes the form
\ba
T_{\cA \cB} (\tilde\phi , \phi)
 &=& e^{-(d_\cA+d_\cB)\phi}\,\sum_{j=-2\ell}^{2\ell} \;
\tilde{T}^{(j)}_{\cA\cB}(\tilde{\phi})
\;,
\ea
where $\tilde{T}^{(j)}_{\cA\cB}$ has charge $(d_\cA+d_\cB- j)$.
At this point we can factor out the dilaton dependence by writing 
the potential \Ref{potential1} in the form
\ba
W(\tilde\phi, \phi) &=& 
\sum_{n=-4\ell}^{4\ell} \; \sum_{k=0}^{4\ell-n} 
\; e^{-(n+k)\phi} \, \tilde{W}^{(n,k)}(\tilde{\phi}) \;,
\label{potgrading}
\ea 
where $\tilde{W}^{(n,k)}$ depends only on $\GTh$-bilinears
$\GTh^{(j_1)}\GTh^{(j_2)}$ with $j_1+j_2 =n$, and has charge
$k$. Moreover, since the potential~\Ref{potential} is obtained from
contracting bilinears in the $T$-tensor $T_{\cA \cB}T_{\cC \cD}$ with
a metric invariant under the compact subgroup of $\mathfrak{g}$, the
components $\tilde{W}^{(n,k)}(\tilde{\phi})$ in \Ref{potgrading}
vanish for $n+k$ odd. After boosting, the potential
becomes
\ba 
\overline{W} &=&
e^{-2\maxl\phi}\,\sum_{k=0}^{2\ell-\maxl} \;
e^{-2k\phi}\,\tilde{W}^{(2\maxl,2k)} \;.
\label{potboost}
\ea 
{}From the form of this potential it is immediately evident that the
boosted potential corresponding to a non-semisimple gauge group in
general will not admit a fully supersymmetric groundstate at $\CV =
{\bf 1}$, even if the original theory did have one, because of the
unbalanced exponential terms.

As an illustration let us consider the theory discussed in
section~\ref{IIAS7} which is obtained from the maximal
four-dimensional gauged supergravity upon reduction on a circle
$S^1$. As discussed above, and in accordance with the
grading~\Ref{IIA}, the scalar content of the three-dimensional theory
comprises the $70$ four-dimensional scalar fields (of charge~$0$), the
$28+28$ contributions from the four-dimensional vector fields (of
charge~$1$), and the two scalars coming from dilaton and graviphoton
(of charge~$0$ and $2$, respectively). With the dilaton defined in
\Ref{dilaton}, the dimensional reduction is performed together with a
Weyl rescaling of the three-dimensional metric $g_{\mu\nu}\rightarrow
e^{-2\phi} g_{\mu\nu}$ in order to obtain a canonical Einstein-Hilbert
term. It is straightforward to verify that the dilaton powers of the
kinetic terms in three dimensions precisely correspond to the
grading~\Ref{IIA}.  Moreover, it is easy to see that the
four-dimensional potential and the kinetic term of the vector fields
in four dimensions give rise to the following scalar terms in three
dimensions
\ba
\sqrt{g_4}\, W_4 &\rightarrow& \sqrt{g_3}\, e^{-2\phi}\, 
\tilde{W}^{(2,0)}  \;,
\non
\sqrt{g_4}\, g^{44} \, A^{ab}_4 A^{cd}_4 \, M_{ab,cd} &\rightarrow&
\sqrt{g_3}\, e^{-4\phi}\, \tilde{W}^{(2,2)} \;,
\ea
(e.g. in the first line we have a factor $e^\phi$ from \Ref{dilaton}
and a factor $e^{-3\phi}$ from the Weyl rescaling, etc.). The above
expressions thus precisely reproduce the first terms of the
expansion~\Ref{potboost}. In this case the series \Ref{potboost} does
not extend to all $2\ell-\maxl+1$ terms due to the fact that the
highest level is a singlet under the gauge group.  Although stationary
points of the four-dimensional potential, i.e.\ of $\tilde{W}^{2,0}$
do not give rise to stationary points of the boosted
potential~\Ref{potboost}, there are indications~\cite{Fisc02,Fisc03}
that they may all be lifted to stationary points of the full
three-dimensional potential~\Ref{potgrading} of the compact gauged
theory. The precise mechanism of the lift remains to be explored; the
series \Ref{potboost} provides a natural starting point, describing
the embedding of the higher-dimensional potential into the
three-dimensional one.

Let us finally mention that the expansion \Ref{potboost} may be
extended to an expansion w.r.t.\ several scalar fields associated with
an abelian subalgebra of $\mathfrak{g}$ using the techniques developed
in \cite{CJLP97}. For particular choices, this corresponds to the 
theories coming from reduction of higher dimensional gauged 
supergravities, with the different terms in \Ref{potboost} 
corresponding to the terms of different higher dimensional origin.

{\small
\paragraph{Acknowledgments}

This work is partly supported by EU contract HPRN-CT-2000-00122 and
HPRN-CT-2000-00131. 
}

\bigskip
\bigskip

{\small 


}

\end{document}